\documentclass[a4paper]{jpconf}
\usepackage{graphicx}
\usepackage{amsmath}
\usepackage{amsfonts}
\usepackage{amssymb}
\usepackage{multirow}

\newcommand{\mc}{\mathcal}
\newcommand{\nl}{\newline}
\newcommand{\be}{\begin{eqnarray}}
 \newcommand{\ee}{\end{eqnarray}}
 \newcommand{\nee}{\nonumber\end{eqnarray}}
 \newcommand{\nn}{\nonumber\\}

  \newcommand{\bc}{\begin{center}}
 \newcommand{\ec}{\end{center}}
 \newcommand{\pup}{p^\uparrow}

\def\kt{k_\perp}

\def\avk{\langle k_\perp ^2\rangle}
\def\avp{\langle p_\perp ^2\rangle}
\def\avPT{\langle P_T^2\rangle}
\def\xb{x_{_{\!B}}}

\def\J{_{_J}}

\def\BM{_{_{B\!M}}}

\def\s              {\sigma}

\begin{document}
\title{ Tests  for the extraction of  Boer-Mulders
functions}\footnote{Talk give at XVII Advanced Research Workshop on
High Energy Physics (DSPIN-17), JINR-Dubna, Russia}

\author{\underline{Ekaterina Christova}}

\address{Institute for Nuclear Research and Nuclear Energy, Bulgarian Academy of Sciences, Sofia, Bulgaria}

\ead{echristo@inrne.bas.bg}

\author{Elliot Leader}

\address{Imperial College, London, UK}

\ead{e.leader@imperial.ac.uk}

\author{Michail Stoilov}

\address{Institute for Nuclear Research and Nuclear Energy, Bulgarian Academy of Sciences, Sofia, Bulgaria}

\ead{mstoilov@inrne.bas.bg}

\begin{abstract}
At present, the Boer-Mulders (BM) functions are extracted from asymmetry data
using the simplifying assumption of their proportionality to the
Sivers functions for each quark flavour.   Here we present two
independent tests for this assumption.
  We subject COMPASS  data on semi-inclusive deep inelastic scattering
 on the $\langle\cos\phi_h\rangle$,  $\langle\cos 2
\phi_h\rangle$ and Sivers  asymmetries to these tests.
  Our analysis   shows that the tests are satisfied  with
    the available data if the proportionality constant is the same for all quark flavours, which does not correspond to
 the flavour dependence used  in existing analyses.     This
     suggests that the  published information on the BM functions
      may be unreliable.

The $\langle\cos\phi_h\rangle$ and   $\langle\cos 2 \phi_h\rangle$
asymmetries receive contributions also from the, in principle,
calculable Cahn effect. We succeed in extracting the Cahn
contributions  from experiment (we believe for the first time) and
compare with their calculated values, with interesting
implications.

\end{abstract}

\section{Introduction}

At present it is already recognized that the collinear picture of the parton model,
 according to which  quark momenta are parallel to  proton momentum,
 is a rather rough approximation for the nucleon structure -- quarks have also transverse momentum.
 This leads especially to a completely new type -- T-odd,  parton densities (pdf's). Here we shall focus on two of them --
 the Boer-Mulders (BM)  and Sivers transverse momentum dependent (TMD) parton densities.

In present analyses \cite{BM_2} it is assumed that the BM functions
are proportional to the Sivers functions. This much simplifies
the analysis, but
 it is clearly  model dependent -- a different assumption would lead to  different BM functions.

 Here we suggest two independent tests for  the above assumption using only measurable quantities -- relations between
the $\langle\cos\phi_h\rangle$,  $\langle\cos 2\phi_h\rangle$ and Sivers  asymmetries. We work with the so-called difference asymmetries
  i.e. the difference
  between  the production of particles and their anti-particles.
   We then utilise  COMPASS data on semi-inclusive deep inelastic scattering (SIDIS)
  on a deuteron target in the formulated tests. Further, as  the $\langle\cos\phi_h\rangle$ and  $\langle\cos 2\phi_h\rangle$
   azimuthal asymmetries  receive contributions from both the BM and Cahn effects,
 we are able to  extract information on the Cahn effect from experiment - as far as we know for the first time. The details are presented in \cite{CLS}.

\section{BM and Sivers functions}

The BM parton densities $\Delta^N f_{q^\uparrow\!\!/p}(\xb ,\kt)$
describe the distribution of transversely polarized quarks
$q^\uparrow$ in an unpolarized proton  \cite{BM}. The Sivers parton
densities $\Delta^N f^{Siv}_{q/\pup}(\xb, \kt)$ describe the
distribution of unpolarized quarks
 in a
 transversely polarized proton $\pup$ \cite{Sivers}.

For the BM and Sivers functions, as well as for all TMD functions that
enter our cross sections,  we use the standard parametrization. For
SIDIS on deuterium target  only the sum of the valence quarks, $Q_V\equiv u_V+d_V$, enters
and we have \cite{we}:
 \be
\hspace*{-.5cm}  \Delta  f^{Q_V}_J(\xb,\kt ,Q^2) \!=\! \Delta
f^{Q_V}\J(\xb,Q^2)\; \sqrt{2e}\,\frac{\kt}{M\J} \;
\frac{e^{-\kt^2/\avk\J }}{\pi\avk\J },\qquad J=BM,
Siv\label{BM-Siv_dist1}
\ee with
 \be
 \Delta
f^{Q_V}\J(\xb,Q^2)\!=\! 2\,{\cal N}\J^{Q_V}(\xb)\,Q_V(\xb,Q^2),\qquad \avk \J= \frac{\avk  \, M^2\J}{\avk  + M^2 \J}\cdot
\label{BM-Siv_dist2}
 \ee  %
Here  the ${\cal N}^{Q_V}\J(\xb)$ are unknown functions, and $M\J$,
or equivalently $\avk \J$,
are unknown  parameters.

\section{The transverse quark momenta}\label{TMD-unpol}

Further in our considerations the parameters  $\avk$ and $\avp$, which
appear in the unpolarized cross sections that
normalize all TMD asymmetries  will enter. They
 are interpreted as the average transverse quark and hadron momenta and are determined from multiplicities. At  present the
 obtained values are rather controversial:

$\bullet$ From old measurements we have:\\
 1)  $\avk \approx 0.25\,GeV^2$ and $\avp \approx 0.20\,GeV^2$~\cite{Anselmino_2005}, from the
  EMC  and FNAL  data;
\\
  2) $\avk = 0.18\,GeV^2$ and $\avp = 0.20\,GeV^2$ \cite{MonteCarlo}.

$\bullet$ The  more recent  data from HERMES and COMPASS separately,
gives quite different values \cite{TMD}:
\\
3) $\avk = 0.57 \pm 0.08\,GeV^2$ and $\avp = 0.12\pm 0.01\,GeV^2$,
extracted from  HERMES data
\\
4) $\avk = 0.61 \pm 0.20\,GeV^2$ and $\avp = 0.19\pm 0.02\,GeV^2$,
extracted from  COMPASS data.

  Further we shall  comment on this controversial situation,
 since the Cahn effect,  which we   extract from data,
  is  calculable, and depends sensitively on $\avk$ and $\avp$.


  \section{The difference asymmetries}

We consider the production of charged hadrons $h^\pm$ in SIDIS of charged leptons on an unpolarized and a transversely polarized deuteron target:
\be
l+d\to l'+h^\pm +X,\qquad l+d^\uparrow\to l'+h^\pm +X
\ee
 We work with the so called difference asymmetries:
   \be
  A^{h^+ - h^-}\equiv \frac{\Delta \s^{h^+}-\Delta \s^{ h^-}}{\s^{h^+} -\s^{ h^-}}\cdot\label{A}
\ee
 where $\s^{h^+, \,h^-}$   and $\Delta \s^{h^+, \,h^-}$ are  the  unpolarized and polarized  SIDIS cross sections, respectively.

 The difference asymmetries do not present a new measurement -- they are expressed in terms of the usual  asymmetries $ A^{h^+,\, h^-}$
 and the ratio of the corresponding multiplicities $r$ \cite{COMPASS-diff}:
 \be
    A^{h^+ -h^-} = \frac{1}{1-r} \left( A^{h^+} - r A^{h^-}\right),
    \qquad A^{h^+}= \frac{\Delta \sigma^{h^+}}{\sigma^{h^+}}, \quad A^{h^-}= \frac{\Delta \sigma^{h^-}}{\sigma^{h^-}},
    \quad  r = \frac{\sigma^{h^-}}{\sigma^{h^+}}.
    \label{eq.Diff1}
     \ee

  As shown in ref.\cite{we},  the advantage of using the difference asymmetries is that,
   based only on charge conjugation
(C) and isospin (SU(2)) invariance of the strong interactions,
   they are expressed  purely in terms of the best known valence-quark
    distributions and  fragmentation functions;
 sea-quark and gluon distributions do not enter. For a deuteron target there is
 the additional simplification -- independently of the final hadron,  only the sum of the valence-quark distributions   enters.\\


\section{The  azimuthal asymmetries }

$\Delta f^{BM}$ and  $ \Delta f^{Siv}$ are measured via the  dependence on the azimuthal angle $\phi_h$ of the final hadron:
\be
d\s^h(x, z, Q^2, P_T,\phi)&=& d\s_0^h\,\left\{1+ A^h_{\cos \phi}\,\cos \phi_h+A^h_{\cos 2\phi}\,\cos 2\phi_h+...\right.\nn
&&\left.+S_T\left[ A^h_{Siv}\,\sin (\phi_s-\phi_h)+..\right]\right\}
\ee
where $d\s_0^h$ is the unpolarized, $\phi_h$-independent cross section, $d\s_0^h\propto \,f_q(x)\otimes D_q^h(z)$.

The asymmetries   $ A_{\cos \phi}^h$  \& $ A_{\cos 2\phi}^h$ receive contributions from both
$\Delta f_{BM}$  and the  Cahn   effect --  $A_{\cos \phi}$ receives a subleading $1/Q$-contribution from both BM and Cahn effects:
 \be
1)\,\, A^h_{\cos \phi}\simeq \frac{1}{Q}\,\sum_q
\left[\Delta f_{q^\uparrow\!\!/p}^{BM}\otimes \Delta D_{q\uparrow}^h
 + f_q\otimes D_q^h\right]
 \ee
 while $ A^h_{\cos 2\phi}$ receives a leading BM-contribution and twist-4 $1/Q^2$-contribution from Cahn effect:
 \be
2)\,\,  A^h_{\cos 2\phi}\simeq \,\sum_q \left[\Delta f_{q^\uparrow\!\!/p}^{BM}\otimes
\Delta D_{q\uparrow}^h
 +\frac{1}{Q^2}\, f_q\otimes D_q^h\right]
\ee
The Sivers asymmetry $A_{Siv}$ is induced by $\Delta f^{Siv}_q$:
\be
3) \,\, A^h_{Siv}\simeq \,\sum_q \Delta f^{Siv}_q\otimes D_q^h
 \ee

\section{Tests for the relation between the BM and Sivers functions on a deuteron target}\label{Relations 1}

As  on a deuterium target
 only the sum of the valence-quarks $Q_V$ enters,
 in contrast to the currently used assumption of proportionality between BM and Sivers functions for each quark  flavour,
  we assume the  relation:
\be
 \Delta
f^{Q_V}\BM(x,k_\perp,Q^2)=\lambda_{Q_V}\,\Delta
f^{Q_V}_{Siv}(x,k_\perp,Q^2),\qquad Q_V=u_V+d_V\label{BM1}
 \ee
 where $\lambda_{Q_V}$ is a constant.

 This assumption leads to  relations between
the BM induced contributions in $A^h_{\cos \phi}$ or
$A^h_{\cos 2\phi}$ and the Sivers asymmetries.
 They are particularly simple and present predictive tests for (\ref{BM1})
  when the  $Q^2$-evolution of the collinear pdf's and fragmentation functions
  (FFs)  can be neglected.
    Here we present these tests.

  Taking into account both the BM and Cahn contributions to the unpolarized cross section, the
 relation  (\ref{BM1})
leads to the following relations between the $\xb$-dependent
  $A^h_{\cos \phi}$ or $A^h_{\cos 2\phi}$  and Sivers  asymmetries on deuterium target \cite{CLS}:
\be
A_{\cos\phi ,d}^{h-\bar h}(\xb)-\Phi (\xb)\,C_{\widetilde{BM}}^h\,A_{Siv,d}^{h-\bar h}(\xb )&=& \Phi (\xb)\,
 C_{Cahn}^h,  \label{R1}\\
 A_{\cos 2\phi ,d}^{h-\bar h}(\xb )-\hat \Phi (\xb )\, \hat C_{\widetilde{BM}}^h\,\,  A_{Siv,d}^{h-\bar h}(\xb)
  &=&\frac{MM_d}{\langle Q \rangle^2}\,\hat \Phi (\xb )\,
  \hat C_{Cahn}^h, \quad h=\pi^+, K^+, h^+\label{R2}
\ee
Here the functions $\Phi (\xb)$ and $\hat\Phi (\xb)$ are completely fixed by kinematics:
\be
\Phi (\xb)=\frac{\sqrt{\pi}\,(2-\bar y)\sqrt{1-\bar y}}{\langle Q \rangle\,[1+(1-\bar y)^2]},\quad
\hat \Phi (\xb)=\frac{2\,(1-\bar y)}{[1+(1-\bar
y)^2]},\quad \bar y = \frac{\langle Q \rangle^2}{2M_dE\,x_B},
\label{Phi}
\ee
$\langle Q \rangle^2$ is some mean value of $Q^2$ for each $\xb$-bin, $M_d$ is the mass of the deuterium target.
 $C_{Cahn}^{h},\,\hat C_{Cahn}^{h}$ and $C_{\widetilde{BM}}^{h},\,\hat C_{\widetilde{BM}}^{h}$ are constants, determined entirely by the collinear
 and Collins FFs, the explicit expressions are given in ref.\cite{CLS}.

Relations  (\ref{R1}) and (\ref{R2}), in which  $C_i^h$, respectively $\hat C_i^h$ are  parameters,
  present:

   1) two independent direct tests of the assumed relation (\ref{BM1}) between the BM and
Sivers functions,  in which only  measurable quantities   enter, and
no knowledge about the TMD functions is required and,

2) two independent ways for  extracting the Cahn contribution from
data.
  \nl

\section{Tests using the COMPASS data for $h^\pm$ production on a deuterium target}

Here we use  relations (\ref{R1}) and (\ref{R2}) between the difference asymmetries   to test  assumption (\ref{BM1})
between the TMD functions,  using the COMPASS
  data on deuterons for production of charged hadrons $h^\pm$ for the  angular distributions
  $A_{\cos \phi,d}^{h^\pm}(\xb )$ and $A_{\cos 2\phi,d}^{h^\pm}(\xb )$
  \cite{COMPASS-UU}, and the single-spin  Sivers asymmetry data $A_{Siv,d}^{h^\pm}(\xb )$  \cite{COMPASS_Siv}.  We proceed in 3 steps:\nl

\begin{figure}[h]
 \centering
\includegraphics[scale=.7]{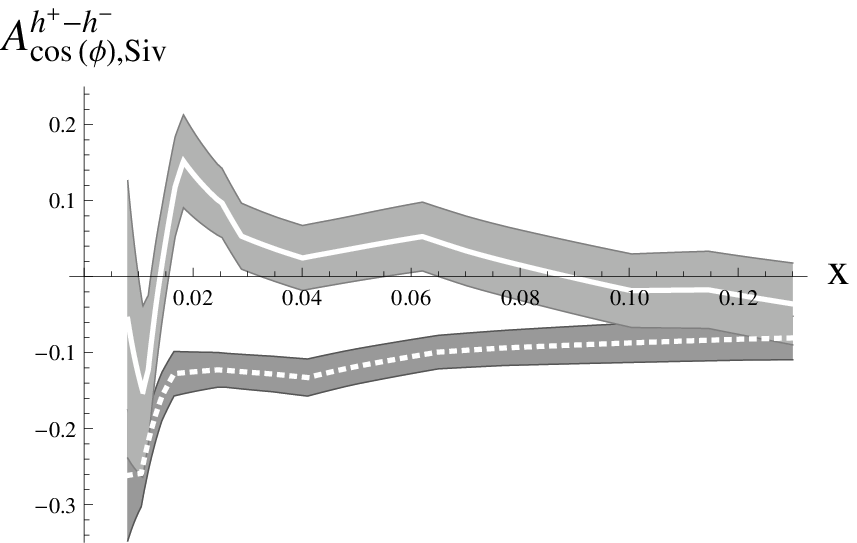}\;\;\includegraphics[scale=.7]{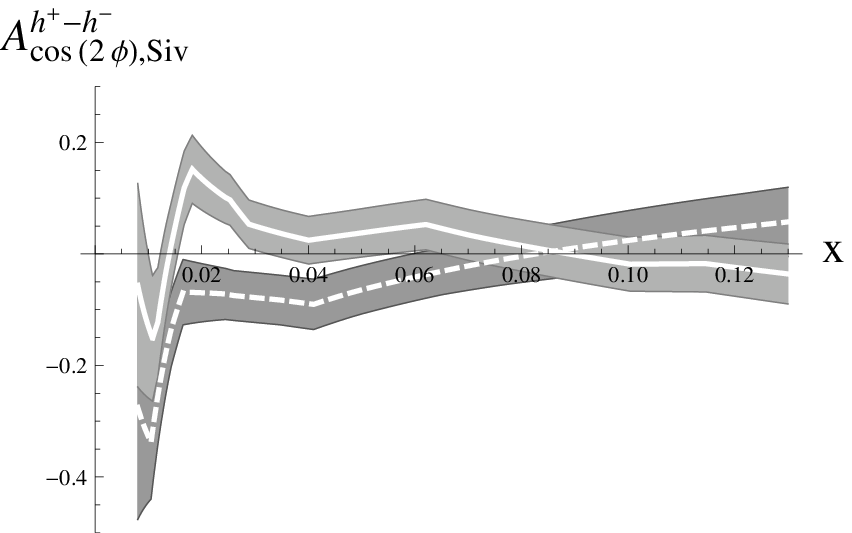}\\
a\hspace{6cm}b
\caption{ The difference asymmetries with their statistical errors : a) $A_{Siv,d}^{h^+ - h^-}(\xb )$  (solid line)
 and $A_{\cos\phi,d}^{h^+ -h^-}(\xb)$ (dashed line)
  b)  $A_{Siv,d}^{h^+ - h^-}(\xb )$  (solid line)
 and $A_{\cos 2\phi,d}^{h^+ -h^-}(\xb)$ (dashed line) }\label{data}
\end{figure}

1) Via Eq.(\ref{eq.Diff1}),  we  form the difference  asymmetries
 $A^{h^+-h^-}_J,\,J=\cos\phi , \cos 2\phi, Siv$
 from the measured  asymmetries $A_j^{h^+}$ and $A_j^{h^-}$ for
positive  and negative charged hadron production,
 $r$   is given in \cite{COMPASS-diff}.
The obtained  asymmetries are presented on Fig.\ref{data}. As seen from the Figure,  the Sivers asymmetry
$A_{Siv,d}^{h-\bar h}(\xb )$ is   determined with
large relative errors and is close to 0.
This suggests that
$C^h_{\widetilde{BM}}$ and $\hat C^h_{\widetilde{BM}}$  may  be poorly determined.
The same  arguments hold for  $A_{\cos 2\phi,d}^{h^+ -h^-}(\xb)$ and we expect
tests with $A_{\cos\phi,d}^{h^+ -h^-}(\xb)$ to give more precise information.

2) We choose the $Q^2$ interval where the  $Q^2$-dependence of the collinear pdf's and FFs can be neglected.
In the COMPASS kinematics
 to each value of $\langle Q^2\rangle$ corresponds one definite value of $\langle \xb \rangle$,
  thus fixing the $Q^2$ interval we
 fix also the $\xb$-interval.
Using the available CTEQ parametrizations for the pdf's and   the parametrization
 in~\cite{LSS-13} for FFs we see that  aside from the small values of $Q^2 < \,1.5\, GeV^2$, the $Q^2$-dependence is weak.
This implies that Eqs~(\ref{R1},\ref{R2}) can be applied for $\xb\geq 0.014$.

3) Finally, we fit the parameters in Eqs~(\ref{R1},\ref{R2}) from
the data, using $\chi^2$-analysis with linear interpolation of the
data. There are two ways to utilize (\ref{R1}) and (\ref{R2}), we
shall follow both of them: \nl

\begin{figure}[htb]
\begin{center}
\includegraphics[scale=.6]{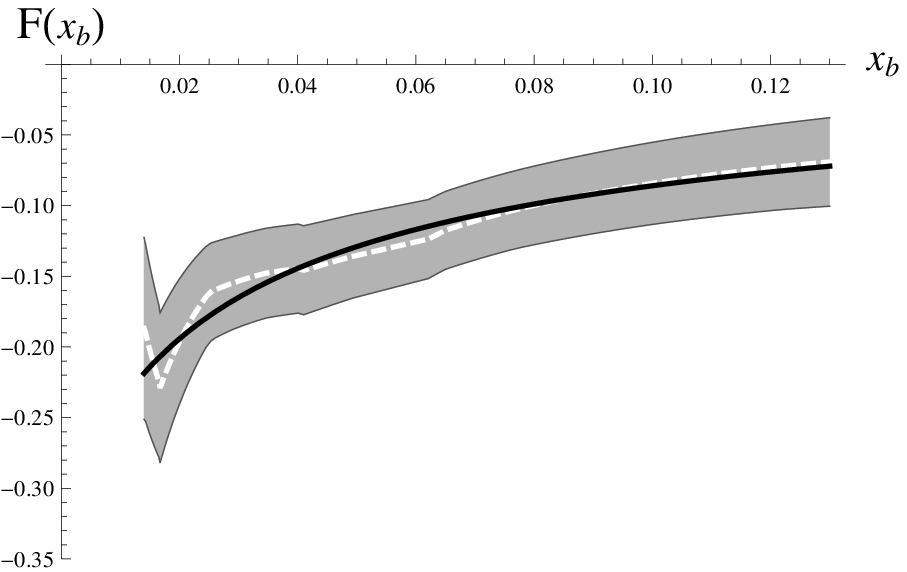}\;\;\;\;\includegraphics[scale=.6]{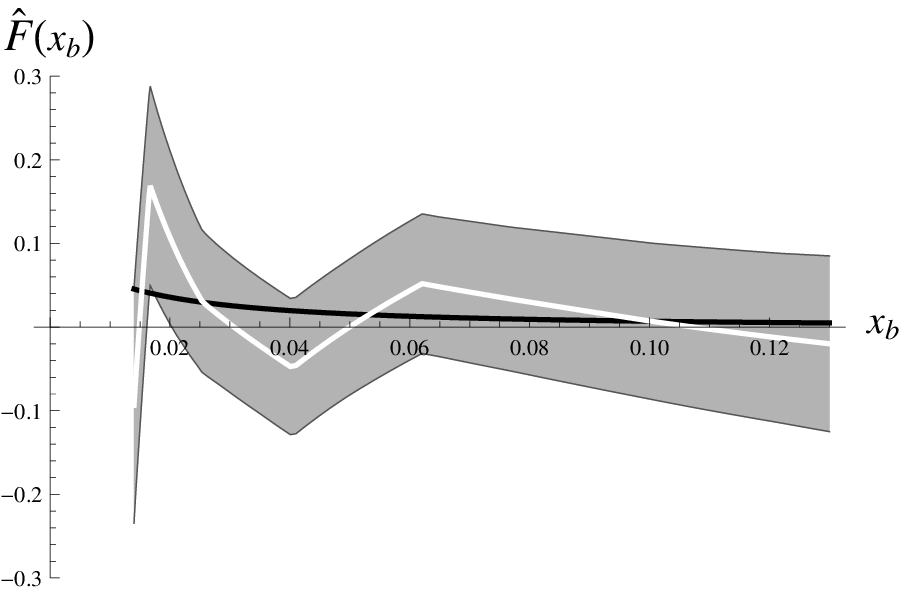}\\
a\hspace{6cm}b
\end{center}
\caption{ The fits for $\xb \geq 0.014$: a) of Eq.(\ref{R1})),
the dashed white line is for $F(\xb )\equiv  A_{\cos\phi,d}^{h^+-h^-}(\xb)-
C_{\widetilde{BM}}^h\,\Phi (\xb)A_{Siv,d}^{h-\bar h}(\xb )$, the black line is $C_{Cahn}^h\,\Phi (\xb)$;
  b)  of Eq.(\ref{R2})), the  white line is $\hat F(\xb)\equiv   A_{\cos 2\phi,d}^{h^+-h^-}(\xb )
- \hat C^h_{\widetilde{BM}}\,\hat \Phi (\xb )\,  A_{Siv,d}^{h^+-h^-}(\xb)$,
  the black one is $MM_d\,\hat\Phi (\xb )\hat C_{Cahn}^h\,/\langle Q \rangle^2 $ } \label{fitco}
\end{figure}

($\mc{{A}}$) We consider  both  $ C_{Cahn}^{h}$ and $ C_{\widetilde{BM}}^{h}$
(respectively $ \hat C_{Cahn}^{h}$ and $ \hat
C_{\widetilde{BM}}^{h}$),  as fitted parameters.  Our analysis showed that for both tests,  Eqs~(\ref{R1},\ref{R2}),
 we obtain a good fit in almost
the same kinematic interval $\xb \gtrsim 0.014$,
 the results   are presented on Fig.{\ref{fitco}}.

($\mc{{B}}$) In the second approach, first we calculate the Cahn constants, $ C_{Cahn}^{h}$ or $\hat C_{Cahn}^{h}$,
 using their explicit expressions. For example, for $ C_{Cahn}^{h}$ we have:
 \be
  C_{Cahn}^h&=&-\avk\, \frac{\int dz_h\,z_h[D_{q_V}^{h}(z_h)]/\sqrt{\avPT}}{\int dz_h\, [D_{q_V}^{h}(z_h)]},\quad
  \avPT =\avp +z_h^2 \avk \label{Cahn}
  \ee
  They
depend   on
  the FFs  and on  the parameters
 $\avk$ and $\avp$ which, as  discussed in  Section (\ref{TMD-unpol}),   vary considerably.
 Then we
  fit  the same data  with just one parameter $ C_{\widetilde{BM}}^{h}$, respectively  $ \hat C_{\widetilde{BM}}^{h}$.
 Consequently, the main interest in this  approach is to compare the calculated Cahn constants
  with  the parameters as in ($\mc{{A}}$).\nl

The  calculated values for $C_{Cahn}^h$ and $\hat C_{Cahn}^h$ for the different values of  $\avk$, $\avp$, and  the fitted parameter
    $C_{\widetilde{BM}}^h$ and $\hat C^h_{\widetilde{BM}}$ are presented in Tables~\ref{fit2}. For comparison, in the last column  the corresponding
fitted values from approach $\mc{{A}}$ are also presented. The  errors in the parameters are calculated using Monte Carlo simulation.

\begin{table}[h]
\centering
\begin{tabular}{|c|c|c|c|c|c|}
  \hline
  &&&&\\
$\avk\; [\mathrm{GeV}^2]$ &
 0.25 & 0.18 & $ 0.57 \pm 0.08$ & $0.61 \pm 0.20$ & ($\mc{A}$)\\
$\avp\; [\mathrm{GeV}^2]$ &
 0.20 & 0.20 & $0.12 \pm 0.01$ & $0.19 \pm 0.02$ & \\
\hline
  &&&&\\
  $C_{Cahn}^h$&
 -0.21&
 -0.16 &
 $-0.49 \pm 0.05$ &
 $-0.4 \pm 0.1$ & $-0.167\pm 0.043 $\\
  \hline
    &&&&\\
 $C^h_{\widetilde{BM}}$
 & 1.43 & 0.44 & $13 \pm 2$ & $11 \pm 4$ & $0.55\pm 0.88 $    \\
 \hline
  &&&&\\
  $\hat{C}_{Cahn}^h$&
 0.077 &
 0.044 &
$ 0.41 \pm 0.08 $&
$ 0.32 \pm 0.15$ & $0.083\pm 0.22 $\\
  \hline
    &&&&\\
$\hat{C}^h_{\widetilde{BM}}$
 & -1.86 & -1.59 & $-4.9 \pm 0.8 $& $-4.0 \pm 1.5 $ & $-1.6\pm 1.6 $ \\
 \hline
\end{tabular}
\caption{$C_{Cahn}^h$ and   $\hat{C}_{Cahn}^h$ are  calculated using  FFs  from  LSS~\cite{LSS-13},
$C_{\widetilde{BM}}^h$ and $\hat{C}^h_{\widetilde{BM}}$ are
fitted. Their values are compared to those of (${\cal A}$).} \label{fit2}
\end{table}

To the best of our knowledge this is the first time that the Cahn
contribution  has been determined from data
 and it is
intriguing that its value is in agreement with a calculated value
based on  the early values of the Gaussian parameters
$\avk= 0.18,\; \mathrm{GeV}^2$ and $\avk= 0.25,\;  \mathrm{GeV}^2$  and completely disagree with the later values.


\section{Conclusions}

We have performed two independent tests of the
 assumption that the  proportionality between the BM and Sivers functions holds
 for the sum of the valence-quark TMD distributions,  using the COMPASS  data  on the  asymmetries
$A_{\cos\phi,d}^{h^+ -h^-}(\xb)$,  $A_{\cos 2\phi ,d}^{h^+
-h^-}(\xb)$
 and $A_{Siv,d}^{h^+ - h^-}(\xb )$. Both tests are consistent with this assumption
  in the same  kinematic interval $\xb = [0.014 , 0.13]$.

 However, in the  published extractions of the   BM functions  \cite{BM_2} the assumption made is
 $\Delta f^q \BM = \lambda_q \Delta f^q_{Siv}$ for each quark separately.
 It  would agree with our result only if
$\lambda_u = \lambda_{\bar u}=\lambda_d = \lambda_{\bar d} =
\lambda_{Q_V}$, which does not correspond to
 the values obtained in \cite{BM_2}.   This suggests that the published  BM functions
 may be unreliable.

We have also determined the kinematical Cahn contribution, both
directly from a fit to the data  (as far as we know for the first
time) and from a calculation.
  The calculated values are very sensitive to  the average transverse momentum-squared $\avk$. Surprisingly,
   the calculated values
 agree with the extracted ones
only for  the old experimental values
 $\avk\approx 0.18\;  \mathrm{GeV}^2$
 and $\avk\approx 0.25\;  \mathrm{GeV}^2$  and completely disagree with the much bigger present-day values.

\section*{Acknowledgments}

 E.Ch. and M.S. acknowledge the support of Grant D-08-17  of the Bulgarian Science Foundation.
 E. L. is grateful to the  Leverhulme Trust for an Emeritus Fellowship and E.Ch. to a collaborative Grant with JINR, Dubna.

\section*{References}


 \end{document}